\documentclass[twocolumn]{revtex4}

\usepackage{xcolor}
\usepackage{amsmath, amsfonts, amssymb, graphicx, color}
\newcommand{\beq}{\begin{equation}}
\newcommand{\eeq}{\end{equation}}

\begin{document}

\title{A quantitative theory of viral--immune coevolution is within reach}

\author{Thierry Mora}
\affiliation{Laboratoire de physique de l'\'Ecole normale sup\'erieure,
	CNRS, PSL University, Sorbonne Universit\'e, and Universit\'e Paris-Cit\'e,
	75005 Paris, France}
\author{Aleksandra M. Walczak}
\affiliation{Laboratoire de physique de l'\'Ecole normale sup\'erieure,
	CNRS, PSL University, Sorbonne Universit\'e,  and Universit\'e Paris-Cit\'e,
	75005 Paris, France}

\begin{abstract}
Pathogens drive changes in host immune systems that in turn exert pressure for pathogens to evolve. Quantifying and understanding this constant co-evolutionary process has clear practical global health implications. Yet its relatively easier accessibility compared to macroevolution makes it a fascinating system to learn about the basic laws of evolution. Focusing on immune--viral evolution, we present an overview of theoretical and experimental approaches that have recently started coming together to build the foundations for a quantitative and predictive co-evolutionary theory.

\end{abstract}

\maketitle

Vertebrates, including us, use their adaptive immune system to protect themselves against recurring or endemic pathogens such as influenza and SARS-CoV-2, or to battle chronic infections such as HIV. Understanding and quantitatively predicting when and how pathogens evolve and escape immunity through escape mutations is of great importance for the design of vaccine strategies, diagnostic tools, and treatment.
The evolution of pathogens is partly driven by the immune system of the hosts. This happens on several time and spatial scales, from within-host evolution in chronic diseases, to continuous interaction with many hosts at the population level in acute infections. The immune systems, in turn, update their memory repertoires to protect hosts against future infections by the pathogen or its close variants. This leads to a process of coevolution between immune systems and pathogens which spans many scales, from the molecular and cellular interactions between immune receptors and pathogenic epitopes, to the organismal and epidemiological levels (Fig.~1).

\begin{figure*}
\begin{center}
  \includegraphics[width=.8\textwidth]{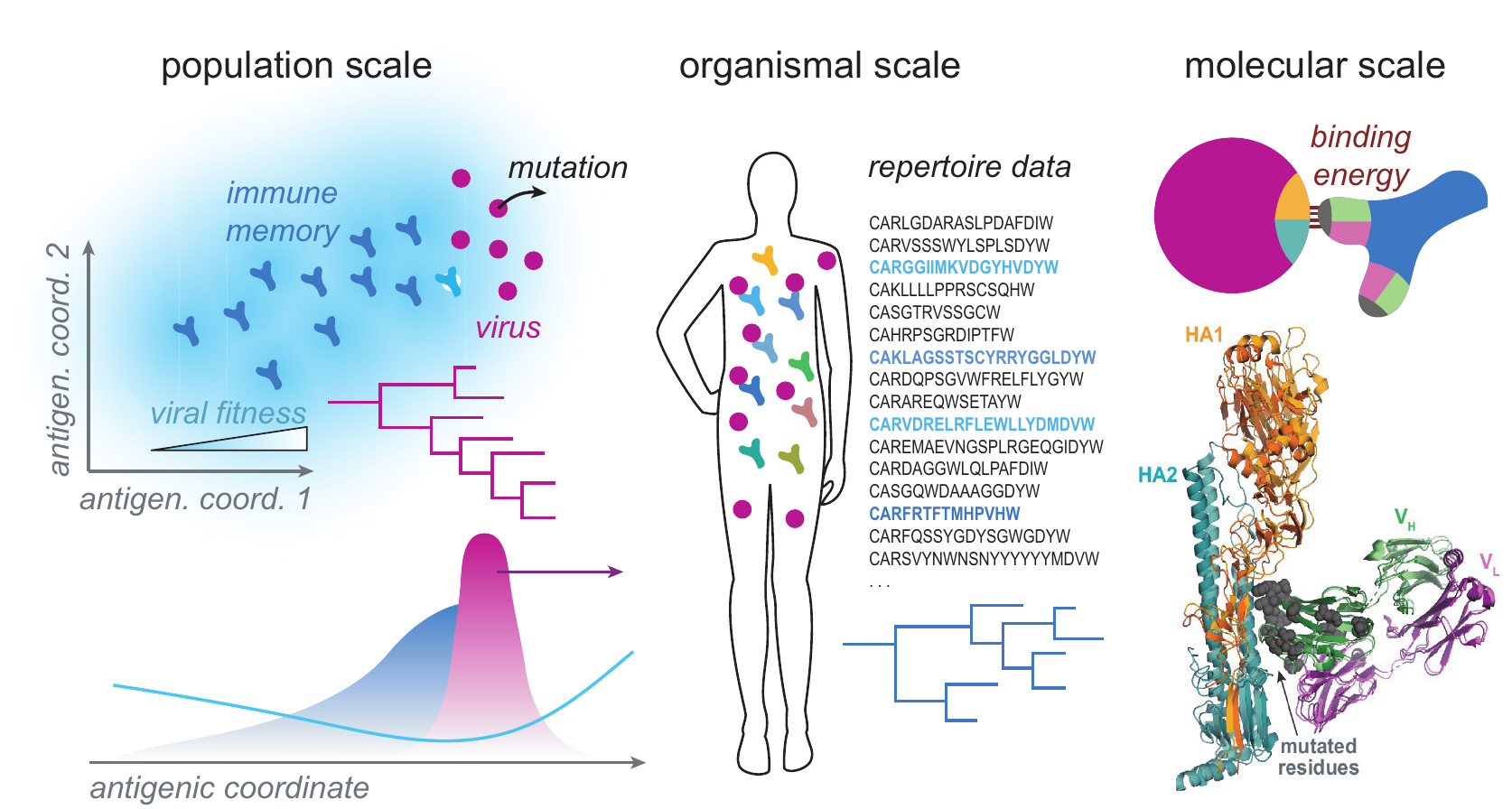}
\end{center}
\caption{{\bf Coevolution from the population to the molecular scale.} Recognition starts at the molecular level through the specific binding of antibodies to bits of the viral proteins called epitopes, illustrated on the right by the co-crystal structure of an antibody fragment and the hemagglutinin protein of influenza (from \cite{Phillips2021b}). At the other end of the population scale (on the left), effective descriptions of antibodies and viruses in a common antigenic space explain how viruses (in purple) are pushed to mutate continually under the pressure of the immune system of its target hosts (in blue), growing faster in regions of the antigenic space where the population is susceptible. At the intermediate scale of the organism (middle), the repertoire of antibodies, represented here by a list of sequences of their third heavy-chain Complementarity Determining Loop, evolves under the selective pressure of successive acute infections (or of a chronic infection) to optimize protection and to prepare the organism for future challenges. Only a small fraction of the whole repertoire (shown in blue) is specific to the viral epitope, and is structured into distinct lineages.} 
\label{fig:Figure1}
\end{figure*}

Protection against pathogens involve all branches of the immune system that work together. The ensemble of antibodies is one branch that has the ability to somatically mutate and adapt rapidly, making it an important element of coevolution occurring on fast timescales. Antibodies presented on the surface of B cells bind to bits of molecules called antigens and recognize them as either natural to the organism (self-proteins) or foreign, triggering other arms of the immune system. Recognition of foreign antigens also induces migration of these B cells to germinal centers, where they somatically hypermutate and are selected for binding to this specific antigen. Along this process hypermutated B cells are released to the blood stream and secrete soluble versions of their antibody that bind to pathogens and help combat the infection more rapidly. This results in a constantly changing, diverse set of expressed antibodies, termed the repertoire.

Viral--antibody coevolution, besides its practical importance for health, offers a testbed for basic questions of evolution and coevolution. As is becoming clear in many fields, organisms do not evolve in steady conditions, competing only with members of their own species. The outcome of evolution is shaped by environmental conditions, resource availability and surrounding species. Viral--immune evolution occurs on fast timescales, with both the viral population and B cell repertoires updating their diversity within weeks. The rapidity of the process, combined with relative ease of sampling compared to other systems, and in some cases the well defined perimeter where the interactions take place, allows us to test and develop quantitative evolutionary theories.

Quantitative evolutionary theories have been around for over 100 years, epidemiological models of susceptible-infected and recovered (SIR) populations close to 100 years~\cite{Kermack1927}, and theories of host--pathogen coevolution date back at least 30 years \cite{Sasaki1994}. In the recent years, in particular in the wake of the COVID-19 pandemic, troves of genomic data and ever more powerful parallelized biophysical assays have offered new opportunities to test these theories experimentally, and to refine quantitative models. We describe how these advances in theoretical models, experimental and population data have all contributed to our quantitative understanding of coevolutionary processes. Yet, taken separately, these amazing feats of technology and imagination often fall short of providing a predictive theory of evolution. While evolution is clearly driven by rare events, coevolution does not allow for all possible subsequent events and may significantly limit actual evolutionary trajectories. Making concrete statistical predictions about future dominant strains and estimating the errors and timescales for the reliability of these predictions is now within reach. Bringing together data, quantitative controlled experiments and theory we can change our understanding of antibody-viral coevolution and propose a predictive quantitative theory of coevolution. 

\section*{The viral--antibody coevolutionary process}
Viruses follow basic Darwinian evolutionary rules. They mutate randomly, and their reproductive success depends on their ability to survive in hosts or to infect new ones. Viral capsid envelope and other externally displayed proteins (such as the spike protein in SARS-CoV-2) breach the host's protective layers, allowing the virus to hijack host cells to reproduce. For this reason, host immune systems target these proteins. Upon succesful recognition of a viral protein as a foreign protein, B cells are recruited to germinal centers where the receptor version of their antibody get hypermutated and selected for better binding to this specific antigen~\cite{Victora2012,Chakraborty2017}. The affinity maturation process goes through a few rounds of mutation and selection~\cite{Victora2012,Tas2016,Weisel2017}, leading to long-lived plasma cells producing highly efficient neutralizing antibodies. At the same time, B cells are also released into the periphery at intermediate stages as memory B cells \cite{Akkaya2020}. As a result, the outcome of affinity maturation is a set of B cells with a range of binding affinities~\cite{Smith1997}, and not just the best binders for this specific antigen.

These antibodies are also cross-reactive, meaning that they can recognize many similar antigenic strains of the virus. Both cross-reactivity and the diversity of antibodies encoded by memory B cells guarantees broad protection against the viral strain as well as many of its antigenic variants. For this reason, the virus needs to accumulate enough mutations on its antigenic sites to avoid recognition by the immune systems of previously infected hosts. In the case of acute infections, it needs these escape mutations to be able to re-infect individuals. Without these escape mutations, it can and does infect susceptible individuals, but after some time a sufficient fraction of the population develops protecting antibodies (herd immunity), making escape mutations necessary for the virus to continue spreading. A similar process occurs for chronic infections such as HIV, but within a single host, and on a faster time scale. The speed of these antigenic substitutions depends on the mutation rate of the virus, as well as on the strength of selection exerted by the hosts' immune systems. 

\section*{Phenomenological theories}
Most theories of immune-viral coevolution start with dynamical models of susceptible, infected, and recovered (SIR) compartments of hosts~\cite{Kermack1927}, repeated for each antigenic strain. The most basic version of this model considers strains in a one-dimensional discrete arrangement called the stepping-stone model, where viral strains can mutate into their neighbors along that single antigenic dimension~\cite{Sasaki1994}, with possible cross-reactivity (also called cross-immunity) protecting recovered individuals against neighboring strains~\cite{Gog2002}.

In their linearized, deterministic version, these models admit traveling wave solutions along the antigenic axis that fall into the same class as the Fisher-Kolmogorov–Petrovsky–Piskunov (Fisher-KPP) equation of reaction diffusion~\cite{Fisher1937}. Viruses at the front the wave enjoy a susceptible host population, giving them a growth advantage over strains in the bulk of the wave, which suffer from herd immunity. As they infect more individuals, these strains become less and less fit, and are in turn outcompeted by more advanced variants along the antigenic axis. This leads to a non-equilibrium steady state where the average fitness of the viral population remains constant, in a striking realization of the ``Red Queen'' hypothesis, where the fitness benefits obtained by escape mutations are continually compensated by building herd immunity.
Adding cross-immunity may yield non-linear solutions of oscillating waves, leading to clustered viral populations and periods of quiescence \cite{Gog2002,Koelle2006}.

Since these early works, other theories of immune-viral coevolution~\cite{Rouzine2018,Yan2019,Marchi2021} have assumed that viral evolution operates in a different regime, where small-number (demographic) noise at the most advanced tip of the wave drives its motion, giving a striking example of how macroscopic behaviour can be determined by the statistics of rare events \cite{Tsimring1996,Desai2007,Rouzine2008,Brunet2008,Good2012}.
The crossover between this stochastic regime and the Fisher-KPP regime is governed by the range of cross-reactivity \cite{Chardes2023}.

The two regimes lead to very different predictions for the rate
of adaptation and its dependence upon the antigenic mutation rate and host population size.
In the regime of small cross-reactivity, the virus spreads in an essentially susceptible population, and the wave falls in the Fisher-KPP regime. The stochastic regime is achieved at large cross-reactivities, very small fitness differences lead to a large advantage to ``jumping before the crowd'' and infecting new hosts. The two regimes correspond to distinct predictions for the evolution of non-antigenic traits such as the mutation rate itself, or the virulence of the pathogen. In the stochastic regime, viruses strive to maximize their basic reproductive number, while in the Fisher-KPP regime they tend to maximize the speed of the wave. Rough estimates seem to place the evolution of influenza and possibly SARS-CoV-2 in the stochastic regime \cite{Chardes2023}.

A limit of these models is the assumption that the antigenic space is one-dimensional.
Dating back to the work of A. Perelson and collaborators~\cite{Segel1989}, physicists have proposed models of coupled non-linear equations in high dimensional space to describe the effective ecology of repertoires, and to predict their response to immune challenges. Within this description antigens and antibodies co-exist in the same space and their ability to interact depends on their mutual distance. The cross-reactivity range defines the distance within which an antibody can recognize an antigen. In this view, antibodies ``cover'' the space of antigens with cross-reactivity hyperspheres.
Hemagglutination inhibition (HI) assays performed in the mid-2000s~\cite{Smith2004} measured the ability of different animal serra  to neutralize influenza viruses found in human populations across the second part of the 20th century. The neutralization trajectory could be computationnally mapped onto a low dimensional trajectory, justifying the approximation made by many phenotypic models.

Models of antigenic waves were extended to multiple or even infinite dimensions (infinite allele model), leading to mathematical developments with direct biological relevance, in particular about the predictability of viral evolution. The theory predicts that the immune pressure pushes the viral wave forward, channeling its inertial course into an effectively one-dimensional track \cite{Bedford2012}, but the direction of this evolutionary trace diffuses rotationally in antigenic space \cite{Marchi2019,Marchi2021}. This observation may be used to retrospectively explain and justify the assumption that antigenic evolution is low-dimensional. However, the rotational diffusion limits the ability to predict future strains, especially in high dimensions.
Accounting for the finite size of the population of infected hosts leads to the possibility that the viral population goes to extinction, or splits into two antigenically independent lineages \cite{Yan2019}. The latter happens when two viral strains at the tip of the wave jump away from the immune systems, but in different directions, and far enough to be outside the cross-reactivity range of each other. This effect, which is reminiscent of the split between the Yamagata and Victoria strain of influenza B in 1983~\cite{Rota1990}, can only occur in multiple dimensions, but is predicted to be a very rare and essentially unpredictable event, with exponential sensitivity to the model parameters \cite{Yan2019,Marchi2021}. The Red Queen state of constant escape is not asymptotically stable, but it is an incredibly long-lived transient. This result proposes a more subtle interpretation of the neutralization data, where the emergence of an effective low dimensional coordinate for viral escape is not guaranteed but results from the parameters of the coevolutionary process.

\section*{Population level viral data}

What does the evolution of real viruses look like? Clearly the impact of viral evolution on our lives is more complex than predicted by theoretical models described above, due to geographic effects~\cite{Dalziel2018, Bedford2010,Bedford2015,Wen2016}, seasons~\cite{Petrova2018}, immune history~\cite{Lewnard2018,Dugan2020, Tsang2023} and social behaviour~\cite{Luca2018, Balcan2009}. Tracking viral evolution based on sampled influenza, HIV and SARS-CoV-2 data has become an important endavour for both global health and policy making~\cite{Perelson1996, Korber2001,Pullano2021,Bubar2021,Ferretti2020a}, as well as understanding coevolution~\cite{Illingworth2014,Meijers2022}.

Since viruses evolve continuously by acquiring unique sets of mutations, these mutations can be used as markers to track evolution and disease spreading. Tracking viral genomes and reconstructing their phylogenies allows us identify how viral populations spread in host populations and how and where epidemics arise.  Bedford and Neher developed the comprehensive NextStrain open source project ({\url{ https://nextstrain.org/}}) that continually updates publicly available pathogen genome data from many species, and offers analysis and interpretation tools~\cite{Hadfield2018}. Collecting data and making it easily accessible is a necessary step for understanding evolutionary processes.

Previously, Neher {\em et al.} used tree branching statistics to succesfully predict the future success of influenza strains~\cite{Neher2014}. In parallel efforts, \L{}uksza  and L\"assig used fitness models based on physical quantities such as protein stability and antigenic binding to predict the evolution of influenza~\cite{Luksza2014}. Such approaches are now used by global health agencies to inform influenza vaccine design. These two models focus on different effective signatures of viral evolution: Neher et al~\cite{Neher2014} exploit signatures of natural selection encoded in the branching structure of viral lineages; \L{}uksza  and L\"assig~\cite{Luksza2014} integrate information about viral stability and antigenicity to build a predictive fitness model. Incidentally, similar fitness models generalize well to a different immune context where coevolution plays a similar conceptual role: the arms race between tumour cells and the T cells that eliminate them. These models are able to predict the immunogenicity of neoantigens~\cite{Luksza2017}, and to detect the selection or elimination of immunogenic cancer cells by the immune system (immuno-editing) \cite{Luksza2022}. They are now being used in the design of patient-specific vaccines for pancreatic cancer~\cite{Rojas2023}. Similarly, fitness models have been used to show that selection induced by vaccination has driven the antigenic evolution of SARS-CoV-2~\cite{Meijers2022}, or to predict the vulnerability of HIV variants with implications for vaccine design \cite{Louie2018}, even leading to a clinical trial \cite{Murakowski2021}.
Despite these important applications,  much work needs to be done to relate these models to more precise descriptions of pathogen-immune interactions, and to integrate them into a general theory of pathogen--immune coevolution.

\section*{Immune repertoire analysis}
On the immune side, recent advances in the profiling of antibody repertoires by high-throughput sequencing \cite{Georgiou2014} have opened opportunities for quantitative understanding of pathogen-immune interactions and coevolution.
In particular, how the immune repertoire is reshaped following infections or re-infections by the same or mutated pathogens is a crucial building block towards building multi-scale models of the coevolutionary arms race between an ensemble of immune systems (from the host population), and a population of evolving pathogens.

The initial diversity of the repertoire is generated by a random process of recombination of the antibody genes, which has been extensively characterized using data-driven biophysical models of recombination and function selection~\cite{Marcou2018,Isacchini2021}. The second process of antibody diversification is through biased hypermutations \cite{Yaari2013,Spisak2020}, which are generated and selected for better binding during the affinity maturation process. Hypermutations create a large number of lineages with distinct ancestors, and each lineage is shaped by selection through their ability to recognize its cognate antigen \cite{Victora2012}. Understanding and quantifying the underlying evolutionary process within antibody repertoires is still a major challenge. Normative or descriptive theories have been proposed to understand how antibody repertoires self-organize to improve their specificity to the target through affinity maturation \cite{Kepler1993a,Meyer-Hermann2012,Wang2015a}, and efficiently encode the memory of past infections to prepare for the next one \cite{Schnaack2021a,Chardes2022}. However, it has been hard to directly relate these prediction to actual repertoire data.

Disease-specific patient cohorts are now routinely subject to antibody repertoire sequencing~\cite{Hoehn2015,Horns2019, Hoehn2019,Roskin2020,Kreer2020a,Nielsen2020a}, with the goal to provide insight into how immunity is shaped and acquired. However, most of these experiments sample whole antibody repertoires from patient blood, which is a mixture of many lineages stemming from multiple past infections. To correctly describe the evolutionary process and gain interpretability, the bulk antibody repertoire data must first be partitioned into clonal families that share common ancestors, and organized into phylogenetic trees retracing the ordering of  hypermutations. Because of the large initial diversity of possible ancestors, lineage clustering is a difficult problem on which recent progress has been made~\cite{Ralph2016,Nouri2020,Spisak2022a}.
In addition, the specific structure of antibody receptor sequences, with varying lengths, small lineage sample sizes due to limited sampling, and heterogeneous mutation rates, makes the problem of phylogenetic reconstruction different and more difficult than in traditional genetics.

Attempts to characterize and quantify the selection on antibody lineages have been inspired by traditional methods from molecular evolution, either by inspecting the spectrum of observed substitutions~\cite{Yaari2012,McCoy2015a}, by analyzing statistics of the lineage structure and phylogenies~\cite{Yaari2015,Horns2019}, or by exploiting longitudinal data to quantify evolution dynamically~\cite{Hoehn2015,Horns2019,Hoehn2019}. 
These analyses have returned global signatures of selection that are statistically significant but hard to interpret, calling for new approaches.

For example, typical measures of selection such as skewed site frequency spectra, abnormal ratios of synonymous to nonsynonymous mutations, or tree imbalance, are detected even in the repertoire of healthy individuals \cite{Horns2019,Nourmohammad2019}, since these repertoires also contain lineages that have been expanded during past infections. This suggests to refine our null hypotheses of what makes a normal healthy repertoire, beyond the standard neutral model of evolution. A detailed analysis~\cite{Spisak2022a} of B cell repertoire datasets from healthy people~\cite{Briney2019} show that evolutionary properties of the lineages do not depend on the global properties of the ancestral sequence, such as its junctional length or generation probability, suggesting that the selection process in germinal centers treats every sequence in the same way. Estimated selection pressures (through ratios of synononymous to nonsynonymous mutations along the trees) span two orders of magnitude, suggesting a broad range of selection forces that go beyond simply purging weak binders (purifying selection). These results based on human data are in agreement with direct observation of diversity in mouse lineages~\cite{Tas2016}, where the binding properties of cells exiting germinal centers is large.

Coming back to the analysis of repertoires in infected individuals, progress has been made possible through the longidutinal analysis of lineages across multiple time points. This has allowed to detect \cite{Horns2019} and subsequently validate \cite{Horns2020} antigen-specific lineages, or to demonstrate the accumulation of hypermutations with time in acute and chronic infections \cite{Hoehn2021}.
Combining longitudinal data with phylogenetic analyses has also helped go beyond traditional genetics approaches, by using a propagator-based approach originally designed to study in viral evolution \cite{Strelkowa2012}. Applied to repertoire data from HIV patients, this method identified which part of the antibody sequence was selected~\cite{Nourmohammad2019}, and demonstated the existence of clonal interference within antibody lineages. In this regime of clonal interference, the best antibodies have a selective advantage driving the growth of their relative frequencies, but are constantly challenged by new variants before they have the time to invade the population. These results reveal a complex dynamics with a continual production of ever better antibodies competing with each other, maintaining a large clonal diversity and potentially slowing down adaptation. 

Despite these advances, repertoire analyses are generally hampered by the fact that the antigenic target of antibodies are mostly unknown, making their practical use limited.
New experimental techniques and computational tools are needed to identify responding or antigen-specific lineages from repertoire data, using either longidutinal data \cite{Horns2020} or large cohors of donors infected with a common disease \cite{RuizOrtega2023}.

\section*{Coevolution}
A recently published HIV dataset was the first to track in depth both antibody repertoires and viral strains in the same patients~\cite{Strauli2019}. While initial analysis does not seem to reveal any signatures of coevolution, pulling the data from different patients and statistically correcting for different sampling depths shows significant anti-correlation between viral and antibody turnover in all patients~\cite{Mazzolini2023}. This suggests that fast viral evolution is accompanied by a slowdown in the change of the composition of the immune repertoire. This is then followed by a fast immune change when the viral population is stable. The results may appear surprising, since we expect the immune system to track the virus, so that their evolutions should be correlated. However, simple string models such as those proposed earlier for HIV evolution \cite{Nourmohammad2016} (see below)
show that both this initial expectation and the observed anti-correlations are correct, depending on the sampling time scale: if it is of the order of the switching time, as is in the experiments, we observe negative correlations, since both the viral and immune populations experience the largest selection pressure, hence the largest turnover,  when the other population is stably adapted, whereas for longer sampling timescales we would observe positive correlations. 

Despite this exception and the weak coevolutionary signal that could be extracted from it, we lack more longitudinal datasets of repertoires and infecting viruses over long time scales. Such experiments would allow us to test coevolution theories quantitatively at the genetic level, with the hope to relate specific viral mutations to specific changes of the repertoire.

Coevolution involves the molecular interaction of viral proteins with immune receptors. While for many purposes a phenomenological description suffices, the molecular nature of these interactions matters, since certain regions of viral proteins are under stronger stability constraints, making them more conserved and more stable targets for the immune system. Conserved regions are usually harder to access from a conformational stand-point and antibodies primarily target strain-specific and variable regions of the virus~\cite{Julien2013}. However, antibodies that do neutralize conserved regions are likely to neutralize more than one viral strain. They are known as broadly neutralizing antibodies (BnAbs), and emerge as a result of natural affinity maturation~\cite{Liao2013}. Yet this happens rarely and usually after long periods of coevolution such as in the case in individuals infected with HIV.  

String models have been introduced to describe the viral protein--antibody interface as strings of interacting amino--acids. Despite their simplicity, these models have been successfully used to understand viral-antibody coevolution \cite{Nourmohammad2016}, and to explain how broadly neutralizing antibodies emerge during affinity selection~\cite{Wang2015, Luo2015}. Using time-shifted neutralization assays in HIV-infected patients~\cite{Richman2003}, fixation of a lineage was shown to be determined by competition between circulating antibody lineages, and BnAbs were found to be more likely to emerge when confronted with diverse viral populations. String models have also helped design vaccination schedules that force affinity maturation into a regime where antibody evolution is focused on the conserved region of the virus (rather than its antigenic sites, which the virus can mutate without much harm), leading to the generation of antibodies with broader affinity~\cite{Wang2015,Wang2017}.

\section*{Immune--antigen interaction at the molecular level}

Quantifying  immune-pathogen  interaction at the molecular level remains an open challenge in understanding their coevolution. The  general problem is to predict binding affinity, or possibly a more functional readout such as antibody neutralization, from the pair of sequences of the antibody on the one hand, and of the pathogenic protein on the other hand. 

Many experiments exist to get insight into the molecular aspects of antibody-antigen interactions in a high-throughput manner, as recently reviewed in~\cite{Moulana2023}:
titration-based binding assays based on yeast display (Tite-Seq), which give the binding affinity of antibody--antigen pairs in the physical units of a dissociation constant ($K_d$)
\cite{Adams2016}, 
neutralization assays~\cite{Dadonaite2023}, epitope specific immunoprecipitation assays~\cite{Xu2015}, hemagglutination assays~\cite{Smith2004}, and directed evolution experiments~\cite{Boder2000}.
Currently these methods allow for testing either one antibody against many antigens, or one antigen against many antibodies. A lot of effort is currently being put into large high-throughput neutralization and binding affinity assays that would allow us to scan many antibody--antigen pairs at a time.

While these methods may seem similar, they probe different phenotypic properties, such as neutralization potency or binding affinity. They also have different scopes: directed evolution experiments aim to find the best binding antibodies~\cite{Boder2000}, or to predict the future course of viral evolution~\cite{Zahradnik2021}. Yet affinity maturation is known to keep not just the best binders~\cite{Tas2016}, and tracing evolutionary trajectories while keeping antibody diversity is essential for describing how coevolution works.
High-throughput assays such as Tite-Seq measure the phenotypic properties of a large library of variants. The choice of the library itself may be different across methods: one may want to assay all possible single mutation variants of the virus (deep mutational scan) to get a complete but local map of mutational effects \cite{Starr2020a}, or explore all possible intermediates between an initial and an evolved variant, either of the antibody \cite{Phillips2021b} or viral \cite{Moulana2022c} sequence.
Combining global and local exploration is essential to map out the constantly changing coevolutionary landscape.

The application of Tite-Seq to libraries of anti-influenza antibody variants has
highlighted the effects of epistatic (non additive) interactions between mutations at many orders, and quantified the constraints that epistasis puts on the evolution of broadly neutralizing antibodies \cite{Phillips2021b}.
Applied to influenza and SARS-CoV-2 evolution, deep mutational scans based on Tite-Seq have shown that the receptor-binding domain (RBD) of the spike protein can support many mutations while maintaining its affinity for ACE2~\cite{Starr2020a}, and have been used to assess monoclonal antibodies for their robustness against antigenic escape, as measured by their affinity to all single mutants of RBD \cite{Starr2021a}. Antibody binding measurements of all possible intermediates between the Wuhan strain and Omicron BA.1 have further revealed the diversity of escape strategies available to the virus, as well as the importance of compensatory mutations \cite{Moulana2022c}.

While the limiting factor for  quantifying antibody--antigen evolution has been mostly experimental, computational scientists have been developing approaches to generalize existing measurements to predict new interacting pairs. The traditional approach based on protein co-structures~\cite{Wilson1990} has proven difficult to generalize into a high-thoughput assay, although as more datasets become available~\cite{Dunbar2014} machine learning approaches are showing potential for generalization. Obtaining a structure does not directly give information about binding affinity, let alone neutralization, and many models aim to bypass the structural step. Some of these efforts have been powered by machine learning techniques~\cite{Shan2022} but simpler sequence based biophysical models of residue preferences and pairwise interactions, such as the string models described above, offer ways of integrating sequencing datasets that are easier to obtain. These models are often trained on deep mutational scan data~\cite{Taft2022}. Since the experiment itself offers only local exploration it is proving hard to learn models that generalize to completely unseen antibodies or antigens.

Mutations first influence binding, however protection is better related to neutalization \cite{Dadonaite2023}. Understanding the coevolutionary landscape beyond simple directed evolution experiments and linking binding and neutralization assays is needed to predictively describe antigen-antibody coevolution. The results of these assays are huge datasets and there is a strong need for theory to make sense of them and to integrate them into predictive models of immune-pathogen coevolution, grounded in their biophysical interactions.

\section*{Conclusions}
The coevolution of immune repertoires and viruses proceed on a complex, rugged binding-affinity landscape, with potential barriers limiting the availability of evolutionary paths.
From the point of view of either a single viral strain, or a single antibody, this landscape is constantly changing as the other component is evolving, creating a dynamic ``seascape'' \cite{Mustonen2009}.
Both the viral and antibody binding landscapes can be mapped out, at least locally, thanks to a combination of molecular evolution binding and neutralization assays and by sampling coevolving immune repertoires and viruses in individuals. Quantitatively predicting and understanding the evolution of both the immune and viral components, and learning how to manipulate this coevolution, could help design vaccines with a broad cross-reactivity (e.g. universal influenza or coronavirus vaccines, or HIV vaccines).

Evolution happens by small molecular changes over long times, leading to new functional solutions. These steps are mutations, insertions or deletions of nucleotides or segments of nucleotides. Most modifications are deleterious to function and most of these are selected against. As a result, not all paths allowed by local evolutionary moves can be implemented in an evolving organism.
For example, for a binding site with $L$ amino-acids, not all $19\times L$ amino acid changes will be viable for the next mutational step.
The fact that evolution is really coevolution between molecules, viral proteins and immune receptors, further restricts the space of realized moves. We are at a point where, by combining data, laboratory experiments and theory we can try to estimate the space of possible evolutionary trajectories and figure out what are the right measures in which to quantify this space. This does not mean we will be able to deterministically foresee the future, but we can put bounds on the scale and timescale of our uncertainty. This theory should give estimates about the diversity and overlap of circulating strains as a function of time (with confidence intervals),  predict the probability of immune specific immune repertoires recognizing a given strain,  and put a likelihood on how surprised we should be to see a very diverse strain appear supposedly ``out of nowhere.'' 

Current phenomenological models of coevolution give us predictions for how stable the coevolutionary co-existence of the viral and host populations is. From the practical public health perspective, a transient state that persists for millions of generations can be viewed as stable. One important remaining challenge is to figure out in what regime particular viruses are, to be able to assess the likelihood that they split into distinct substrains, or eventually go extinct. In order to do that, we need to better calibrate these phenomenological models using data---both population data sampled from individuals, and molecular data from evolutionary experiments. The goal of having a quantitative falsifiable theory of coevolution is within our reach.

\bibliographystyle{pnas}

\begin{thebibliography}{100}

\bibitem{Phillips2021b}
Phillips AM, {et~al.}
\newblock (2021) Binding affinity landscapes constrain the evolution of broadly
  neutralizing anti-influenza antibodies.
\newblock \emph{eLife} 10:e71393.

\bibitem{Kermack1927}
Kermack W, McKendrick A
\newblock (1927) Contributions to the mathematical theory of epidemics.
\newblock \emph{Bltn Mathcal Biology} 53:33--55.

\bibitem{Sasaki1994}
Sasaki A
\newblock (1994) Evolution of {{Antigen Drift}}/{{Switching}}: {{Continuously
  Evading Pathogens}}.
\newblock \emph{Journal of Theoretical Biology} 168:291--308.

\bibitem{Victora2012}
Victora GD, Nussenzweig MC
\newblock (2012) Germinal {{Centers}}.
\newblock \emph{Annual Review of Immunology} 30:429--457.

\bibitem{Chakraborty2017}
Chakraborty AK
\newblock (2017) A {{Perspective}} on the {{Role}} of {{Computational Models}}
  in {{Immunology}}.
\newblock \emph{Annual Review of Immunology} 35:403--439.

\bibitem{Tas2016}
Tas JMJ, {et~al.}
\newblock (2016) Visualizing antibody affinity maturation in germinal centers.
\newblock \emph{Science} 351:1048--1054.

\bibitem{Weisel2017}
Weisel F, Shlomchik M
\newblock (2017) Memory {{B Cells}} of {{Mice}} and {{Humans}}.
\newblock \emph{Annual Review of Immunology} 35:255--284.

\bibitem{Akkaya2020}
Akkaya M, Kwak K, Pierce SK
\newblock (2020) B cell memory: Building two walls of protection against
  pathogens.
\newblock \emph{Nature Reviews Immunology} 20:229--238.

\bibitem{Smith1997}
Smith KGC, Light A, Nossal GJV, Tarlinton DM
\newblock (1997) The extent of affinity maturation differs between the memory
  and antibody-forming cell compartments in the primary immune response.
\newblock \emph{The EMBO Journal} 16:2996--3006.

\bibitem{Gog2002}
Gog JR, Grenfell BT
\newblock (2002) Dynamics and selection of many-strain pathogens.
\newblock \emph{Proceedings of the National Academy of Sciences}
  99:17209--17214.

\bibitem{Fisher1937}
Fisher RA
\newblock (1937) The wave of advance of advantageous genes.
\newblock \emph{Annals of Eugenics} 7:355--369.

\bibitem{Koelle2006}
Koelle K, Cobey S, Grenfell B, Pascual M
\newblock (2006) Epochal {{Evolution Shapes}} the {{Phylodynamics}} of
  {{Interpandemic Influenza A}} ({{H3N2}}) in {{Humans}}.
\newblock \emph{Science} 314:1898--1903.

\bibitem{Rouzine2018}
Rouzine IM, Rozhnova G
\newblock (2018) Antigenic evolution of viruses in host populations.
\newblock \emph{PLOS Pathogens} 14:e1007291.

\bibitem{Yan2019}
Yan L, Neher RA, Shraiman BI
\newblock (2019) Phylodynamic theory of persistence, extinction and speciation
  of rapidly adapting pathogens.
\newblock \emph{eLife} 8:e44205.

\bibitem{Marchi2021}
Marchi J, L{\"a}ssig M, Walczak AM, Mora T
\newblock (2021) Antigenic waves of virus\textendash immune coevolution.
\newblock \emph{Proceedings of the National Academy of Sciences}
  118:e2103398118.

\bibitem{Tsimring1996}
Tsimring LS, Levine H, Kessler DA
\newblock (1996) {{RNA Virus Evolution}} via a {{Fitness-Space Model}}.
\newblock \emph{Physical Review Letters} 76:4440--4443.

\bibitem{Desai2007}
Desai MM, Fisher DS, Murray AW
\newblock (2007) The {{Speed}} of {{Evolution}} and {{Maintenance}} of
  {{Variation}} in {{Asexual Populations}}.
\newblock \emph{Current Biology} 17:385--394.

\bibitem{Rouzine2008}
Rouzine IM, Brunet {\'E}, Wilke CO
\newblock (2008) The traveling-wave approach to asexual evolution: {{Muller}}'s
  ratchet and speed of adaptation.
\newblock \emph{Theoretical Population Biology} 73:24--46.

\bibitem{Brunet2008}
Brunet {\'E}, Rouzine IM, Wilke CO
\newblock (2008) The {{Stochastic Edge}} in {{Adaptive Evolution}}.
\newblock \emph{Genetics} 179:603--620.

\bibitem{Good2012}
Good BH, Rouzine IM, Balick DJ, Hallatschek O, Desai MM
\newblock (2012) Distribution of fixed beneficial mutations and the rate of
  adaptation in asexual populations.
\newblock \emph{Proceedings of the National Academy of Sciences}
  109:4950--4955.

\bibitem{Chardes2023}
Chard{\`e}s V, Mazzolini A, Mora T, Walczak AM
\newblock (2023) Evolutionary stability of antigenically escaping viruses.

\bibitem{Segel1989}
Segel LA, Perelson AS
\newblock (1989) Shape space: An approach to the evaluation of cross-reactivity
  effects, stability and controllability in the immune system.
\newblock \emph{Immunology Letters} 22:91--99.

\bibitem{Smith2004}
Smith DJ
\newblock (2004) Mapping the {{Antigenic}} and {{Genetic Evolution}} of
  {{Influenza Virus}}.
\newblock \emph{Science} 305:371--376.

\bibitem{Bedford2012}
Bedford T, Rambaut A, Pascual M
\newblock (2012) Canalization of the evolutionary trajectory of the human
  influenza virus.
\newblock \emph{BMC Biology} 10:38.

\bibitem{Marchi2019}
Marchi J, {et~al.}
\newblock (2019) Size and structure of the sequence space of repeat proteins.
\newblock \emph{PLOS Computational Biology} 15:e1007282.

\bibitem{Rota1990}
Rota PA, {et~al.}
\newblock (1990) Cocirculation of two distinct evolutionary lineages of
  influenza type {{B}} virus since 1983.
\newblock \emph{Virology} 175:59--68.

\bibitem{Dalziel2018}
Dalziel BD, {et~al.}
\newblock (2018) Urbanization and humidity shape the intensity of influenza
  epidemics in {{U}}.{{S}}. cities.
\newblock \emph{Science} 362:75--79.

\bibitem{Bedford2010}
Bedford T, Cobey S, Beerli P, Pascual M
\newblock (2010) Global {{Migration Dynamics Underlie Evolution}} and
  {{Persistence}} of {{Human Influenza A}} ({{H3N2}}).
\newblock \emph{PLOS Pathogens} 6:e1000918.

\bibitem{Bedford2015}
Bedford T, {et~al.}
\newblock (2015) Global circulation patterns of seasonal influenza viruses vary
  with antigenic drift.
\newblock \emph{Nature} 523:217--220.

\bibitem{Wen2016}
Wen F, Bedford T, Cobey S
\newblock (2016) Explaining the geographical origins of seasonal influenza
  {{A}} ({{H3N2}}).
\newblock \emph{Proceedings of the Royal Society B: Biological Sciences}
  283:20161312.

\bibitem{Petrova2018}
Petrova VN, Russell CA
\newblock (2018) The evolution of seasonal influenza viruses.
\newblock \emph{Nature Reviews Microbiology} 16:47--60.

\bibitem{Lewnard2018}
Lewnard JA, Cobey S
\newblock (2018) Immune {{History}} and {{Influenza Vaccine Effectiveness}}.
\newblock \emph{Vaccines} 6:28.

\bibitem{Dugan2020}
Dugan HL, {et~al.}
\newblock (2020) Preexisting immunity shapes distinct antibody landscapes after
  influenza virus infection and vaccination in humans.
\newblock \emph{Science Translational Medicine} 12:eabd3601.

\bibitem{Tsang2023}
Tsang TK, {et~al.}
\newblock (2023) Investigation of the {{Impact}} of {{Childhood Immune
  Imprinting}} on {{Birth Year-Specific Risk}} of {{Clinical Infection During
  Influenza A Virus Epidemics}} in {{Hong Kong}}.
\newblock \emph{The Journal of Infectious Diseases} p jiad009.

\bibitem{Luca2018}
Luca GD, {et~al.}
\newblock (2018) The impact of regular school closure on seasonal influenza
  epidemics: A data-driven spatial transmission model for {{Belgium}}.
\newblock \emph{BMC infectious diseases} 18:29.

\bibitem{Balcan2009}
Balcan D, {et~al.}
\newblock (2009) Multiscale mobility networks and the spatial spreading of
  infectious diseases.
\newblock \emph{Proceedings of the National Academy of Sciences}
  106:21484--21489.

\bibitem{Perelson1996}
Perelson AS, Neumann AU, Markowitz M, Leonard JM, Ho DD
\newblock (1996) {{HIV-1 Dynamics}} in {{Vivo}}: {{Virion Clearance Rate}},
  {{Infected Cell Life-Span}}, and {{Viral Generation Time}}.
\newblock \emph{Science} 271:1582--1586.

\bibitem{Korber2001}
Korber B, {et~al.}
\newblock (2001) Evolutionary and immunological implications of contemporary
  {{HIV-1}} variation.
\newblock \emph{British Medical Bulletin} 58:19--42.

\bibitem{Pullano2021}
Pullano G, {et~al.}
\newblock (2021) Underdetection of cases of {{COVID-19}} in {{France}}
  threatens epidemic control.
\newblock \emph{Nature} 590:134--139.

\bibitem{Bubar2021}
Bubar KM, {et~al.}
\newblock (2021) Model-informed {{COVID-19}} vaccine prioritization strategies
  by age and serostatus.
\newblock \emph{Science} 371:916--921.

\bibitem{Ferretti2020a}
Ferretti L, {et~al.}
\newblock (2020) Quantifying {{SARS-CoV-2}} transmission suggests epidemic
  control with digital contact tracing.
\newblock \emph{Science} 368:eabb6936.

\bibitem{Illingworth2014}
Illingworth CJR, Fischer A, Mustonen V
\newblock (2014) Identifying {{Selection}} in the {{Within-Host Evolution}} of
  {{Influenza Using Viral Sequence Data}}.
\newblock \emph{PLOS Computational Biology} 10:e1003755.

\bibitem{Meijers2022}
Meijers M, Ruchnewitz D, {\L}uksza M, L{\"a}ssig M
\newblock (2022) Vaccination shapes evolutionary trajectories of
  {{SARS-CoV-2}}., ({Biophysics}), Preprint.

\bibitem{Hadfield2018}
Hadfield J, {et~al.}
\newblock (2018) Nextstrain: Real-time tracking of pathogen evolution.
\newblock \emph{Bioinformatics} 34:4121--4123.

\bibitem{Neher2014}
Neher RA, Russell CA, Shraiman BI
\newblock (2014) Predicting evolution from the shape of genealogical trees.
\newblock \emph{eLife} 3.

\bibitem{Luksza2014}
{\L}uksza M, L{\"a}ssig M
\newblock (2014) A predictive fitness model for influenza.
\newblock \emph{Nature} 507:57--61.

\bibitem{Luksza2017}
{\L}uksza M, {et~al.}
\newblock (2017) A neoantigen fitness model predicts tumour response to
  checkpoint blockade immunotherapy.
\newblock \emph{Nature} 551:517--520.

\bibitem{Luksza2022}
{\L}uksza M, {et~al.}
\newblock (2022) Neoantigen quality predicts immunoediting in survivors of
  pancreatic cancer.
\newblock \emph{Nature} 606:389--395.

\bibitem{Rojas2023}
Rojas LA, {et~al.}
\newblock (2023) Personalized {{RNA}} neoantigen vaccines stimulate {{T}} cells
  in pancreatic cancer.
\newblock \emph{Nature} 618:144--150.

\bibitem{Louie2018}
Louie RHY, Kaczorowski KJ, Barton JP, Chakraborty AK, McKay MR
\newblock (2018) Fitness landscape of the human immunodeficiency virus envelope
  protein that is targeted by antibodies.
\newblock \emph{Proceedings of the National Academy of Sciences}
  115:E564--E573.

\bibitem{Murakowski2021}
Murakowski DK, {et~al.}
\newblock (2021) Adenovirus-vectored vaccine containing multidimensionally
  conserved parts of the {{HIV}} proteome is immunogenic in rhesus macaques.
\newblock \emph{Proceedings of the National Academy of Sciences}
  118:e2022496118.

\bibitem{Georgiou2014}
Georgiou G, {et~al.}
\newblock (2014) The promise and challenge of high-throughput sequencing of the
  antibody repertoire.
\newblock \emph{Nature Biotechnology} 32:158--168.

\bibitem{Marcou2018}
Marcou Q, Mora T, Walczak AM
\newblock (2018) {{IGoR}}: A tool for high-throughput immune repertoire
  analysis.
\newblock \emph{Nature Communications} 9:561.

\bibitem{Isacchini2021}
Isacchini G, Walczak AM, Mora T, Nourmohammad A
\newblock (2021) Deep generative selection models of {{T}} and {{B}} cell
  receptor repertoires with {{soNNia}}.
\newblock \emph{Proceedings of the National Academy of Sciences}
  118:e2023141118.

\bibitem{Yaari2013}
Yaari G, {et~al.}
\newblock (2013) Models of {{Somatic Hypermutation Targeting}} and
  {{Substitution Based}} on {{Synonymous Mutations}} from {{High-Throughput
  Immunoglobulin Sequencing Data}}.
\newblock \emph{Frontiers in Immunology} 4.

\bibitem{Spisak2020}
Spisak N, Walczak AM, Mora T
\newblock (2020) Learning the heterogeneous hypermutation landscape of
  immunoglobulins from high-throughput repertoire data.
\newblock \emph{Nucleic Acids Research} 48:10702--10712.

\bibitem{Kepler1993a}
Kepler TB, Perelson AS
\newblock (1993) Cyclic re-entry of germinal center {{B}} cells and the
  efficiency of affinity maturation.
\newblock \emph{Immunology Today} 14:412--415.

\bibitem{Meyer-Hermann2012}
{Meyer-Hermann} M, {et~al.}
\newblock (2012) A {{Theory}} of {{Germinal Center B Cell Selection}},
  {{Division}}, and {{Exit}}.
\newblock \emph{Cell Reports} 2:162--174.

\bibitem{Wang2015a}
Wang S, {et~al.}
\newblock (2015) Manipulating the {{Selection Forces}} during {{Affinity
  Maturation}} to {{Generate Cross-Reactive HIV Antibodies}}.
\newblock \emph{Cell} 160:785--797.

\bibitem{Schnaack2021a}
Schnaack OH, Nourmohammad A
\newblock (2021) Optimal evolutionary decision-making to store immune memory.
\newblock \emph{eLife} 10:e61346.

\bibitem{Chardes2022}
Chard{\`e}s V, Vergassola M, Walczak AM, Mora T
\newblock (2022) Affinity maturation for an optimal balance between long-term
  immune coverage and short-term resource constraints.
\newblock \emph{Proceedings of the National Academy of Sciences}
  119:e2113512119.

\bibitem{Hoehn2015}
Hoehn KB, {et~al.}
\newblock (2015) Dynamics of immunoglobulin sequence diversity in {{HIV-1}}
  infected individuals.
\newblock \emph{Philosophical Transactions of the Royal Society B: Biological
  Sciences} 370:20140241.

\bibitem{Horns2019}
Horns F, Vollmers C, Dekker CL, Quake SR
\newblock (2019) Signatures of selection in the human antibody repertoire:
  {{Selective}} sweeps, competing subclones, and neutral drift.
\newblock \emph{Proceedings of the National Academy of Sciences}
  116:1261--1266.

\bibitem{Hoehn2019}
Hoehn KB, {et~al.}
\newblock (2019) Repertoire-wide phylogenetic models of {{B}} cell molecular
  evolution reveal evolutionary signatures of aging and vaccination.
\newblock \emph{Proceedings of the National Academy of Sciences}
  116:22664--22672.

\bibitem{Roskin2020}
Roskin KM, {et~al.}
\newblock (2020) Aberrant {{B}} cell repertoire selection associated with
  {{HIV}} neutralizing antibody breadth.
\newblock \emph{Nature Immunology} 21:199--209.

\bibitem{Kreer2020a}
Kreer C, {et~al.}
\newblock (2020) Longitudinal {{Isolation}} of {{Potent Near-Germline
  SARS-CoV-2-Neutralizing Antibodies}} from {{COVID-19 Patients}}.
\newblock \emph{Cell} 182:843--854.e12.

\bibitem{Nielsen2020a}
Nielsen SC, {et~al.}
\newblock (2020) Human {{B Cell Clonal Expansion}} and {{Convergent Antibody
  Responses}} to {{SARS-CoV-2}}.
\newblock \emph{Cell Host \& Microbe} 28:516--525.e5.

\bibitem{Ralph2016}
Ralph DK, Iv FAM
\newblock (2016) Likelihood-{{Based Inference}} of {{B Cell Clonal Families}}.
\newblock \emph{PLOS Computational Biology} 12:e1005086.

\bibitem{Nouri2020}
Nouri N, Kleinstein SH
\newblock (2020) Somatic hypermutation analysis for improved identification of
  {{B}} cell clonal families from next-generation sequencing data.
\newblock \emph{PLOS Computational Biology} 16:e1007977.

\bibitem{Spisak2022a}
Spisak N, Dupic T, Mora T, Walczak AM
\newblock (2022) Combining mutation and recombination statistics to infer
  clonal families in antibody repertoires.

\bibitem{Yaari2012}
Yaari G, Uduman M, Kleinstein SH
\newblock (2012) Quantifying selection in high-throughput {{Immunoglobulin}}
  sequencing data sets.
\newblock \emph{Nucleic Acids Research} 40:e134--e134.

\bibitem{McCoy2015a}
McCoy CO, {et~al.}
\newblock (2015) Quantifying evolutionary constraints on {{B-cell}} affinity
  maturation.
\newblock \emph{Philosophical Transactions of the Royal Society B: Biological
  Sciences} 370:20140244.

\bibitem{Yaari2015}
Yaari G, Benichou JIC, Vander~Heiden JA, Kleinstein SH, Louzoun Y
\newblock (2015) The mutation patterns in {{B-cell}} immunoglobulin receptors
  reflect the influence of selection acting at multiple time-scales.
\newblock \emph{Philosophical Transactions of the Royal Society B: Biological
  Sciences} 370:20140242.

\bibitem{Nourmohammad2019}
Nourmohammad A, Otwinowski J, {\L}uksza M, Mora T, Walczak AM
\newblock (2019) Fierce {{Selection}} and {{Interference}} in {{B-Cell
  Repertoire Response}} to {{Chronic HIV-1}}.
\newblock \emph{Molecular Biology and Evolution} 36:2184--2194.

\bibitem{Briney2019}
Briney B, Inderbitzin A, Joyce C, Burton DR
\newblock (2019) Commonality despite exceptional diversity in the baseline
  human antibody repertoire.
\newblock \emph{Nature} 566:393--397.

\bibitem{Horns2020}
Horns F, Dekker CL, Quake SR
\newblock (2020) Memory {{B Cell Activation}}, {{Broad Anti-influenza
  Antibodies}}, and {{Bystander Activation Revealed}} by {{Single-Cell
  Transcriptomics}}.
\newblock \emph{Cell Reports} 30:905--913.e6.

\bibitem{Hoehn2021}
Hoehn KB, {et~al.}
\newblock (2021) Human {{B}} cell lineages associated with germinal centers
  following influenza vaccination are measurably evolving.
\newblock \emph{eLife} 10:e70873.

\bibitem{Strelkowa2012}
Strelkowa N, L{\"a}ssig M
\newblock (2012) Clonal {{Interference}} in the {{Evolution}} of {{Influenza}}.
\newblock \emph{Genetics} 192:671--682.

\bibitem{RuizOrtega2023}
Ruiz~Ortega M, Spisak N, Mora T, Walczak AM
\newblock (2023) Modeling and predicting the overlap of {{B-}} and {{T-cell}}
  receptor repertoires in healthy and {{SARS-CoV-2}} infected individuals.
\newblock \emph{PLOS Genetics} 19:e1010652.

\bibitem{Strauli2019}
Strauli N, {et~al.}
\newblock (2019) The genetic interaction between {{HIV}} and the antibody
  repertoire.

\bibitem{Mazzolini2023}
Mazzolini A, Mora T, Walczak AM
\newblock (2023) Inspecting the interaction between human immunodeficiency
  virus and the immune system through genetic turnover.
\newblock \emph{Philosophical Transactions of the Royal Society B: Biological
  Sciences} 378:20220056.

\bibitem{Nourmohammad2016}
Nourmohammad A, Otwinowski J, Plotkin JB
\newblock (2016) Host-{{Pathogen Coevolution}} and the {{Emergence}} of
  {{Broadly Neutralizing Antibodies}} in {{Chronic Infections}}.
\newblock \emph{PLOS Genetics} 12:e1006171.

\bibitem{Julien2013}
Julien JP, {et~al.}
\newblock (2013) Crystal structure of a soluble cleaved {{HIV-1}} envelope
  trimer.
\newblock \emph{Science} 342:1477--1483.

\bibitem{Liao2013}
Liao HX, {et~al.}
\newblock (2013) Co-evolution of a broadly neutralizing {{HIV-1}} antibody and
  founder virus.
\newblock \emph{Nature} 496:469--476.

\bibitem{Wang2015}
Wang C, {et~al.}
\newblock (2015) B-cell repertoire responses to varicella-zoster vaccination in
  human identical twins.
\newblock \emph{Proceedings of the National Academy of Sciences} 112:500--505.

\bibitem{Luo2015}
Luo S, Perelson AS
\newblock (2015) Competitive exclusion by autologous antibodies can prevent
  broad {{HIV-1}} antibodies from arising.
\newblock \emph{Proceedings of the National Academy of Sciences}
  112:11654--11659.

\bibitem{Richman2003}
Richman DD, Wrin T, Little SJ, Petropoulos CJ
\newblock (2003) Rapid evolution of the neutralizing antibody response to
  {{HIV}} type 1 infection.
\newblock \emph{Proceedings of the National Academy of Sciences}
  100:4144--4149.

\bibitem{Wang2017}
Wang S
\newblock (2017) Optimal {{Sequential Immunization Can Focus Antibody
  Responses}} against {{Diversity Loss}} and {{Distraction}}.
\newblock \emph{PLOS Computational Biology} 13:e1005336.

\bibitem{Moulana2023}
Moulana A, Dupic T, Phillips AM, Desai MM
\newblock (2023) Genotype\textendash phenotype landscapes for immune\textendash
  pathogen coevolution.
\newblock \emph{Trends in Immunology} 44:384--396.

\bibitem{Adams2016}
Adams RM, Mora T, Walczak AM, Kinney JB
\newblock (2016) Measuring the sequence-affinity landscape of antibodies with
  massively parallel titration curves.
\newblock \emph{eLife} 5:e23156.

\bibitem{Dadonaite2023}
Dadonaite B, {et~al.}
\newblock (2023) A pseudovirus system enables deep mutational scanning of the
  full {{SARS-CoV-2}} spike.
\newblock \emph{Cell} 186:1263--1278.e20.

\bibitem{Xu2015}
Xu GJ, {et~al.}
\newblock (2015) Comprehensive serological profiling of human populations using
  a synthetic human virome.
\newblock \emph{Science} 348:aaa0698--aaa0698.

\bibitem{Boder2000}
Boder ET, Midelfort KS, Wittrup KD
\newblock (2000) Directed evolution of antibody fragments with monovalent
  femtomolar antigen-binding affinity.
\newblock \emph{Proceedings of the National Academy of Sciences}
  97:10701--10705.

\bibitem{Zahradnik2021}
Zahradn{\'i}k J, {et~al.}
\newblock (2021) {{SARS-CoV-2}} variant prediction and antiviral drug design
  are enabled by {{RBD}} in vitro evolution.
\newblock \emph{Nature Microbiology} 6:1188--1198.

\bibitem{Starr2020a}
Starr TN, {et~al.}
\newblock (2020) Deep {{Mutational Scanning}} of {{SARS-CoV-2 Receptor Binding
  Domain Reveals Constraints}} on {{Folding}} and {{ACE2 Binding}}.
\newblock \emph{Cell} 182:1295--1310.e20.

\bibitem{Moulana2022c}
Moulana A, {et~al.}
\newblock (2022) Compensatory epistasis maintains {{ACE2}} affinity in
  {{SARS-CoV-2 Omicron BA}}.1.
\newblock \emph{Nature Communications} 13:7011.

\bibitem{Starr2021a}
Starr TN, {et~al.}
\newblock (2021) {{SARS-CoV-2 RBD}} antibodies that maximize breadth and
  resistance to escape.
\newblock \emph{Nature} 597:97--102.

\bibitem{Wilson1990}
Wilson IA, Cox NJ
\newblock (1990) Structural basis of immune recognition of influenza virus
  hemagglutinin.
\newblock \emph{Annual Review of Immunology} 8:737--771.

\bibitem{Dunbar2014}
Dunbar J, {et~al.}
\newblock (2014) {{SAbDab}}: The structural antibody database.
\newblock \emph{Nucleic Acids Research} 42:D1140--D1146.

\bibitem{Shan2022}
Shan S, {et~al.}
\newblock (2022) Deep learning guided optimization of human antibody against
  {{SARS-CoV-2}} variants with broad neutralization.
\newblock \emph{Proceedings of the National Academy of Sciences}
  119:e2122954119.

\bibitem{Taft2022}
Taft JM, {et~al.}
\newblock (2022) Deep mutational learning predicts {{ACE2}} binding and
  antibody escape to combinatorial mutations in the {{SARS-CoV-2}}
  receptor-binding domain.
\newblock \emph{Cell} 185:4008--4022.e14.

\bibitem{Mustonen2009}
Mustonen V, L{\"a}ssig M
\newblock (2009) From fitness landscapes to seascapes: Non-equilibrium dynamics
  of selection and adaptation.
\newblock \emph{Trends in genetics: TIG} 25:111--119.

\end{thebibliography}

\end{document}